\documentstyle[preprint,eqsecnum,aps]{revtex}
\def\qv#1{\buildrel{\scriptstyle{\leftrightarrow}}
   \over{\textstyle{#1}}}

\newcommand{\be}{\begin{equation}}
\newcommand{\ee}{\end{equation}}
\newcommand{\bea}{\begin{eqnarray}}
\newcommand{\eea}{\end{eqnarray}}
\newcommand{\bml}{\begin{mathletters}}
\newcommand{\eml}{\end{mathletters}}
\newcommand{\pa}{\partial}

\newcommand{\bp}{\bar{\psi}}
\newcommand{\p}{\psi}

\newcommand{\dxy}{\delta(\vec{x}-\vec{y})}

\newcommand{\vx}{\vec{x}}
\newcommand{\vy}{\vec{y}}

\newcommand{\g}{\gamma}

\newcommand{\e}{\epsilon}
\newcommand{\ve}{\varepsilon}

\begin{document}
\draft
\title{BFT Embedding of Interacting Second-Class Systems \cite{byline}}
\author{Martin Fleck}
\address{Instituto de F\'{\i}sica, USP  \\ C. P. 66318, 05315-970  -
S\~{a}o Paulo, SP, Brazil.}

\maketitle
\begin{abstract}
The embedding procedure of Batalin, Fradkin, and Tyutin, which allows
to convert a second-class system into a first-class one, is employed to  
convert second-class interacting models. Two cases are considered. One, is the Self-Dual 
model minimally coupled to a Dirac fermion field. The other, the Self-Dual model minimally 
coupled to a charged scalar field. In both cases, they are found equivalent interacting 
Maxwell-Chern-Simons type field theories. These equivalences are pushed beyond the formal 
level, by analysing some tree level probability amplitudes associated to the models. 

\end{abstract}

\newpage
\narrowtext

\section{Introduction}
\label{sec:level1}

It is a well estabilished fact, that the free self-dual (SD) model\cite{TPN1}, a second-class 
field theory, can be viewed as a gauge fixed version of the free, first-class, Maxwel-Chern-Simons 
(MCS) model\cite{Jackiw2}. In a first moment, this equivalence was proved semi-classically 
\cite{Jack1} by relating the classical trajectories of the two models and, subsequently, extended 
to the quantum domain by comparing their Green function generating functionals and the 
corresponding green functions associated with them \cite{Rothe1,Rothe5,Fleck3}.

Although this equivalence holds even in the presence of external sources, the question of what 
happens if coupling 
to other dynamical fields are considered, is only partialy solved. Indeed, in the fermionic 
case, it has been shown \cite{Mal} that the SD model minimally coupled to a Dirac field is 
equivalent to a MCS model coupled non-minimally to a magnetic current and an additional 
Thirring like interaction. On the other hand, the question of what happens, when coupling with complex 
scalar fields are considered, is much more involved and remains open. In this case, the 
conserved current depends explicitly on the basic fields. Therefore, approaches based on 
interpolating lagrangians\cite{Rothe1,Schap1} fail because of a non trivial matter field dependence of the corresponding 
functional determinant. 

This paper is devoted to the study of these two cases. The main strategy 
adopted in this work, is to relate Green function generating functionals associated to different 
models in order to conclude about their equivalence. As we intend to relate second to 
first-class theories, the embedding procedure of 
Batalin, Fradkin and Tyutin (BFT) \cite{bft1,bft2}, which allows us to convert a second-class 
system into a first-class one, is employed. Then, an equivalent first-class version for the 
second-class interacting SD model can be constructed.

We start, in Sec. 2, by characterizing the second-class SD model minimally coupled to Dirac fermions 
on the hamiltonian level. The BFT embedding procedure is then used to generate the corresponding 
first-class counterpart. After verifying the existence of a unitary gauge which render the original 
second-class model back, we implement a particular canonical transformation which enable us to 
construct all the phase-space variables of the first-class theory in terms of those of the 
second-class one. Despite of the nonrenormalizability of the resulting first-class model, the 
equivalence is tested in a tree level calculation, by comparing the probability amplitudes for 
the M\"oller scattering process in the two models.

Simillarly, in Sec. 3, the SD model coupled to a complex scalar field is considered. Once the system 
is characterized in the phase-space, the BFT embedding procedure is applied to generate the 
corresponding first-class counterpart. The existence of a unitary gauge condition secures the 
equivalence of the first-class and the original second-class models. Here too, a particular 
canonical transformation is implemented, which relate all the phase-space variables of the first-class 
theory in terms of those of the old second-class one. Then, we present a MCS type lagrangian 
density which has the same phase-space Green function generating functional realization. As in the 
fermionic case, this equivalence is explored by comparing the probability amplitudes for 
the M\"oller scattering process associated to the models.  

The conclusions are contained in Sec. 4.
\section{The SD model coupled to a Dirac fermion field}
\label{sec:level2}
The classical and quantum dynamics of the SD model minimally coupled to a Dirac fermion field is 
described, in configuration space, by the lagrangian density

\be
\label{100}
{\cal L}^{SD}_F= - \frac{1}{2m} \epsilon^{\mu\nu\rho}(\partial_{\mu} f_{\nu})f_{\rho}
+\frac{1}{2} f^{\mu}f_{\mu}+\frac{i}{2}\bar{\psi}\gamma^{\mu}{\qv{D}}_{\mu}^f\psi 
- M \bar{\psi}\psi\,\,\,,
\ee

\noindent
in which $D_{\mu}^f=\pa_{\mu}-igf_{\mu}$ is the covariant derivative, $m$ and $M$ denote the 
boson and the fermion mass respectively and $g$ is 
the coupling constant\footnote{We use natural units $(c=\hbar=1)$. Our metric is 
$g_{00}=-g_{11}=-g_{22}=1$. The fully antisymmetric tensor $\epsilon^{\mu\nu\rho}$ is 
normalized such that $\epsilon^{012}=1$ and we define $\epsilon^{ij}= \epsilon^{0ij}$. Repeated Greek 
indices sum from $0$ to $2$, while repeated Latin indices from $1$ to $2$. The left-right 
derivative is defined by $\Phi{\qv{D}}\Psi=\Phi D\Psi-(D\Phi)\Psi$.} . On the hamiltonian level, 
the system is completely characteryzed by the primary constraints

\bml
\bea
\label{105}
&&T_0^{(0)}=\pi_0\,\approx\,0\,\,\,, \label{mlett:a105} \\
&&T_i^{(0)}=\pi_i
+\frac{1}{2m} \epsilon_{ij} f^j\approx 0\,\,,i=1,2\,\,\,,
\label{mlett:b105}
\eea
\eml

\noindent
the secondary constraint,

\be
\label{110}
T_3^{(0)}=\frac{1}{m}\left(f^0
-\frac{1}{m}\epsilon_{ij}\partial^if^j\right)+\frac{g}{m}\bp\g^0\p
\approx\,0\,\,\,,
\ee

\noindent
and the canonical hamiltonian

\be
\label{115}
H^{(0)}=\int d^2x\,\left[\frac{1}{2}f^0f^0
+\frac{1}{2}f^if^i-i\bp\g^i(\pa_i-igf_i)\p+M\,\bp\p
-\,m f^0T_3^{(0)}\right]\,\,\,.
\ee

In these equations, $\pi_{\mu}$ are the momentum canonical conjugate to the basic fields $f^{\mu}$ 
and the canonical Poisson bracket structure satisfyed by them imply the second-class nature of the 
constraints. We proceed now, defining partial Dirac $\delta$-brackets with respect to the 
constrains (\ref{mlett:b105}), in the usual manner\cite{Di1}. Within the $\delta$-bracket 
algebra, the constraints (\ref{mlett:b105}) holds as strong identities and we use them to eliminate 
from the game the momenta $\pi_i$. Then, the only non vanishing $\delta$-brackets are

\bml
\bea
\label{120}
\left [f^i(x^0,\vx)\,,\,f^j(x^0,\vy)\right]_{\delta}&=&-m\,\epsilon^{ij}\,\dxy\,\,,
\label{mlett:a120 }\\
\left\{\p_a(x^0,\vx)\,,\,\bp_b(x^0,\vy)\,\right\}_{\delta}& = &-i\,
\g^0_{ab}\,\dxy\,\,\,.\label{mlett:b120 }
\eea
\eml

The BFT embedding procedure is now applyed to the remining second-class system, defined by the 
constraints (\ref{mlett:a105}) and (\ref{110}), the Hamiltonian (\ref{115}) and the $\delta$-bracket 
algebra (\ref{120}). For this purpose, one starts by introducing an additional pair of canonical 
variables ($[\theta,\pi_{\theta}]_{\delta}=1$), one for each second-class constraint. The new constraints and 
the new Hamiltonian are found, afterwards, throgh an interative scheme which, in the present case, 
ends after a finite number of steps. Presently, the BFT conversion procedure yields

\bml
\bea
\label{125}
&&T_0^{(0)} \rightarrow T_1 \,=\,\pi_0\,+\,\theta \,\approx\,0\,\,\,,\label{mlett:a125 }\\
&&T_3^{(0)} \rightarrow T_2 \,=\,\frac{1}{m}
(f^0+\pi_{\theta})-\frac{1}{m^2}\e_{ij}\pa^i f^j\,+\,\frac{g}{m}\,\bp\g^0\p \,\approx\,0\,\,\,,\label{mlett:b125 }\\
&&H^{(0)} \rightarrow  H=\int d^2x\left[\frac{1}{2}(f^0+\pi_{\theta})^2+
\frac{1}{2}(f^i+\pa^i\theta)^2-i\bp\g^i(\pa_i-igf_i)\p+
M\bp\p\right]\,\,\,.\label{mlett:c125}
\eea
\eml

\noindent
One can easily check that the new constraints and the new Hamiltonian are, as
required, strong under involution, i.e.,

\bml
\label{140}
\bea
&&[\,T_A\,,\,T_B\,]_{\delta}\,=\,0\,\,\,,\label{mlett:a140}\\
&&[\,T_A\,,\,H\,]_{\delta}\,=\,0\,\,\,.\label{mlett:b140}
\eea
\eml

\noindent
The converted system is, indeed, first-class and obeys an Abelian involution
algebra.
We construct next the unitarizing Hamiltonian ($H_U$) and the corresponding 
Green function generating functional ($W_{\chi}$)\cite{bft1,bft2}. If we denote by

\be
\label{145}
\Psi\, \equiv \,\int d^2x \left( {\bar{\cal C}}_A\,\chi^A\,-\,{\bar {\cal P}}_A\,\lambda^A\right )
\,\,\,,
\ee

\be
\label{150}
\Omega\, \equiv \int d^2x \left ( {\bar {\pi}}_A\,{\cal P}^A\,+\,T_A\,{\cal C}^A\right )\,\,\,,
\ee

\noindent
the gauge fixing fermion function and the BRST charge, respectively, one has 
that

\be
\label{155}
H_U\,=\,H\,-[\,\Psi\,,\,\Omega\,]_{\delta}\,\,\,.
\ee

\noindent
Here, ${\cal C}^A$ and ${\bar{\cal C}}_A$ are ghost coordinates 
and ${\bar{\cal P}}_A$ and ${\cal P}^A$ their respective canonical 
conjugate momenta. Furthermore, $\lambda^A$ is the Lagrange multiplier 
associated with the constraint $T_A$ and ${\bar{\pi}}_A$ is its 
canonical conjugate momentum. The gauge conditions $\chi^A$ are to be chosen 
such that 

\be
\label{160}
\det {[\, \chi^A\,,\,T_B\, ]_{\delta}}\,\neq\,0\,\,\,.
\ee

\noindent
The corresponding Green function generating functional  

\be
\label{165}
W_{\chi}\,=\,{\cal N}\,\int\,[{\cal D}\sigma]\,\exp(iA_U)\,\,\,,
\ee

\noindent
is writen in terms of the unitarizing action $A_U$

\be
\label{170}
A_U=\int d^3x
\left( \pi_{0}{\dot f}^{0}+\frac{1}{2m}f^i\e_{ij}{\dot f}^j+\pi_{\theta}{\dot {\theta}}+
i\bp\g^0{\dot \p}+{\bar {\pi}}_A
{\dot {\lambda}}^A+{\bar {\cal C}}_A{\dot {\cal P}}^A+{\bar {\cal P}}_A{\dot {\cal C}}^A-
{\cal H}_U \right)\,\,,
\ee

\noindent
and the integration measure $[{\cal D}\sigma]$, involves all the variables appearing in
$A_U$. We proceed by restricting ourselves to consider gauge 
conditions which do not depend upon $\lambda^A$ and/or ${\bar {\pi}}_A$. 
Then, the rescaling $\chi^A \rightarrow \chi^A/\beta$, 
${\bar {\pi}}_A \rightarrow \beta {\bar {\pi}}_A$ and 
${\bar {\cal C}}_A \rightarrow \beta {\bar {\cal C}}_A$ allows, at 
the limit $\beta \rightarrow 0$, to carry out all the integrals
over the ghosts and multiplier variables\cite{FVi2,H1}, with the result

\bea
\label{175}
& &{\cal W}_{\chi}\,=\,{\cal N}\,\int\,[{\cal D}f^0]\,[{\cal D}\pi_0]
[{\cal D}f^{i}]\,[{\cal D}\pi_{\theta}]\,[{\cal D}\theta]\,
[{\cal D}\bp][{\cal D}\p]\det[\chi^A,T_B]_{\delta}\nonumber\\
&\times &
\left(\prod_{A=1}^{2}\delta [\chi^A]\right)
\left(\prod_{A=1}^{2}\delta [T_A]\right)
\exp\left[i\int d^3x
\left( \pi_0{\dot f}^0+\frac{1}{2m}f^i\e_{ij}{\dot f}^{j}+\pi_{\theta}{\dot {\theta}}+i\bp\g^0{\dot {\p}} 
-{\cal H}\right)\right].
\eea

\noindent
Here, ${\cal H}$ is the Hamiltonian density corresponding to the involutive 
Hamiltonian (\ref{mlett:c125}). For completeness, we mention that under an infinitesimal 
supertransformation generated by $\Omega$, the reduced phase space variables present 
in (\ref{175}), change as follows

\bml
\label{180}
\bea
\delta\,f^0\,& \equiv &\,[\,f^0\,,\,\Omega\,]_{\delta}\,\ve\,=\,\ve^1
\,\,\,,\label{mlett:a180}\\
\delta\,f^i\,& \equiv &\,[\,f^i\,,\,\Omega\,]_{\delta}\,\ve\,=\,-\frac{1}{m}\,
\pa^i\ve^2\,\,\,,\label{mlett:b180}\\
\delta\,\theta\,& \equiv &\,[\,\theta\,,\,\Omega\,]_{\delta}\,\ve\,=\,
\frac{1}{m}\ve^2\,\,\,,\label{mlett:c180}\\
\delta\,\pi_{0}\,& \equiv &\,[\,\pi_{0}\,,\,\Omega\,]_{\delta}\,\ve\,
=\,-\frac{1}{m}\,\ve^2\,\,\,.\label{mlett:d180}\\
\delta\,\pi_{\theta}\,& \equiv &\,[\,\pi_{\theta}\,,\,\Omega\,]_{\delta}\,\ve\,
=\,-\,\ve^1\,\,\,,\label{mlett:e180}\\
\delta\,\p\,& \equiv &\,[\,\p\,,\,\Omega\,]_{\delta}\,\ve\,
=\,-\frac{ig}{m}\,\p\,\ve^2\,\,\,,\label{mlett:f180}\\
\delta\,\bp\,& \equiv &\,[\,\bp\,,\,\Omega\,]_{\delta}\,\ve\,
=\,+\frac{ig}{m}\,\bp\,\ve^2\,\,\,,\label{mlett:g180}
\eea
\eml

\noindent
where $\ve^A={\cal C}^A\ve$ and $\ve$ is the BRST infinitesimal constant fermionic 
parameter. These transformation laws show that the first-class theory cannot be enterly 
expressed in terms of the gauge invariant combinations

\bml
\bea
\label{185}
Y^{0}&=&f^0+\pi_{\theta}\,\,,\label{mlett:a185}\\
Y^{i}&=&f^i+\pa^i\theta\,\,.\label{mlett:b185}
\eea
\eml

\noindent
In fact, we observe that, in (\ref{mlett:c125}), an interaction mediated by a gauge 
dependent $f^{i}$ field is what we expect in order to preserve gauge invariance.
Finally, the introduction of the gauge conditions $\chi^1=\pi_{\theta}$ and $\chi^2=\theta$, 
reduces the functional (\ref{175}) to the Green function generating functional 
for second-class systems derived by Senjanovic\cite{Sen}. They represent therefore, the 
unitary gauge conditions.

Most part of the investigations on BFT embedding procedure ends at this stage, by obteining 
the first-class counterpart of a second-class original model. We turn now to look for 
alternative formulations of the first-class theory. For this end, we start by decomposing
the basic $f^i$ field in canonical pairs ($[A^i,P_j]_{\delta}=\delta^i_j$), 

\be
\label{200}
f^i\,=\,\frac{1}{2}A^i\,+\,m\epsilon^{ij}\,P_j\,\,\,,
\ee

\noindent
to write a modified generating functional,

\bea
\label{205}
& &{\cal W}_{\chi}\,=\,{\cal N}\,\int\,[{\cal D}f^0]\,[{\cal D}\pi_0]
[{\cal D}A^{i}][{\cal D}P_{i}]\,[{\cal D}\theta]\,[{\cal D}\pi_{\theta}]\,
[{\cal D}\bp][{\cal D}\p]\det[\chi^A,T_B]_{\delta}\nonumber\\
&\times &
\left(\prod_{A=1}^{2}\delta [\chi^A]\right)
\left(\prod_{A=1}^{2}\delta [T_A]\right)
\exp\left[i\int d^3x
\left( \pi_0{\dot f}^0+P_i{\dot A}^i+\pi_{\theta}{\dot {\theta}}+i\bp\g^0{\dot {\p}} 
-{\cal H}\right)\right].
\eea

\noindent
Notice that, despite the fact that we are left with an essentially canonical phase-space, once 
momentum variables are reintroduced and the particular ${\delta}$-brackets involved 
have a canonical form, the symplectic structure of the constraints (\ref{mlett:b105}) 
is encoded in the composition law (\ref{200}). To show the equivalence of the just mentioned 
generating functional with (\ref{175}), we note that in expression (\ref{205}), the 
Hamiltonian (${\cal H}$) depends on the variables $A^i$ and $P_i$ only through expressions 
(\ref{200}). This suggests the change of variables $A^i \rightarrow A^{\prime i} = f^i$, 
$P_i \rightarrow P^{\prime}_{i} = P_i$, whose jacobian is a nonvanishing real number. Since 
${\cal H}$ does not depend upon $P^{\prime}_{i}$, the corresponding integration is straightforward 
and after performing it one obtains (\ref{175}). With this, we note that in terms of the varables 
$A^i$ and $P_i$, the first-class constraint $T_2$ can be rewriten as

\be
\label{210}
mT_2\,=\,Y^0+{\cal G}-\frac{1}{2m}\epsilon_{ij}F^{ij}+g\bp\g^0\p\,\approx\,0\,\,\,,
\ee

\noindent
where $F^{ij}=\pa^i A^j - \pa^j A^i$ while ${\cal G}$ has the form
 
\be
\label{215}
{\cal G}\,=\,\pa^k P_k\,+\,\frac{1}{2m}\,\e_{kl}\,\pa^k A^l\,\,\,.
\ee

\noindent
We now focus on Eq.(\ref{205}) and perform the particular canonical transformation:

\bml
\bea
\label{220}
&& f^0\,\rightarrow\,A^{\prime 0}\,=\,f^0\,\,\,,\label{mlett:a220}\\
&& \pi_0\,\rightarrow\,P^{\prime}_0\,=\,\pi_0\,+\,\theta \,\approx 0\,\,\,,\label{mlett:b220}\\
&& A^i\,\rightarrow\,A^{\prime i}\,=\,A^i\,\,\,,\label{mlett:c220}\\
&& P_i\,\rightarrow\,P^{\prime}_i\,=\,P_i\,-\,\frac{1}{m}\,\epsilon_{ij}\,\pa^j\theta
\,\,\,,\label{mlett:d220}\\
&& \theta\,\rightarrow\,\theta^{\prime}\,=\,\theta\,\,\,,\label{mlett:e220}\\
&& \pi_{\theta}\,\rightarrow\,\pi_{\theta}^{\prime}\,=\,\pi_{\theta}\,
-\,\frac{1}{2m}\,\e_{ij}\,F^{ij}\,+\,\,f^0\,\,\,,\label{mlett:f220}\\
&& \p\,\rightarrow\,\p^{\prime}\,=\,\p\,\,\,,\label{mlett:g220}\\
&& \bp\,\rightarrow\,\bp^{\prime}\,=\,\bp\,\,\,.\label{mlett:h220}
\eea
\eml

\noindent
Note that the $P^{\prime}_0$ variable is nothing but the first-class constraind $T_1$. Omiting 
the``$\prime$'' superindex in the new variables and performing the $\pi_{\theta}$ integration, 
the generating functional turn to

\bea
\label{225}
{\cal W}_{\chi}&=&{\cal N}\int
[\prod_{\mu=0}^2 {\cal D}A^{\mu}]
[\prod_{\mu=0}^2 {\cal D}P_{\mu}]
[{\cal D}\theta]
[{\cal D}\bp][{\cal D}\p]
\det[\chi^A,T_B]_{\delta}
\left(\prod_{A=1}^{2}\delta [\chi^A]\right)
\delta[P_0]\nonumber\\
&\times &
\exp\left\{i\int d^3x
\left[ P_{\mu}{\dot A}^{\mu}+{\dot {\theta}}{\cal G}+i\bp\g^0{\dot {\p}}-g\bp\g^0\p{\dot {\theta}}
-g\bp\g^i\p\pa_i\theta\right.\right.\nonumber\\
&-&\left.\left.\frac{m^2}{2}P_iP_i+\frac{m}{2}P_i\e^{ij}A^j-\frac{1}{8}A^iA^i
-\frac{1}{4m^2}F^{ij}F^{ij}+{\bp}\left ( i\g^i\pa_i-M\right ){\p}\right.\right.\nonumber\\
&-&\left.\left.\frac{g}{2}\bp\g^i\p A^i-mg\bp\g^i\p\e^{ij}P_j
+\frac{g}{2m}\e_{ij}F^{ij}(\bp\g^0\p)-\frac{g^2}{2}(\bp\g^0\p)^2\right.\right.\nonumber\\
&-&\left.\left.{\cal G}\left(\frac{1}{2}{\cal G}-\frac{1}{2m}\epsilon_{ij}F^{ij}+
g\bp\g^0\p\right)\right]\right\}\,\,,
\eea

\noindent
where we had been made recoursivelly use of the first-class constraints $T_A$. The next step, 
is to define the gauge invariant fermions,

\bml
\bea
\label{230}
\Psi &=&\p\,\exp{\left(-ig\theta\right)}\,\,\,,\label{mlett:a230}\\
\bar{\Psi}&=&\bp\,\exp{\left(+ig\theta\right)}\,\,\,,\label{mlett:b230}
\eea
\eml

\noindent
in terms of which, the generating functional is now writen,

\bea
\label{235}
{\cal W}_{\chi}&=&{\cal N}\,\int\,
[\prod_{\mu=0}^2 {\cal D}A^{\mu}]\,
[\prod_{\mu=0}^2 {\cal D}P_{\mu}]\,
[{\cal D}\theta]\,
[{\cal D}\bar{\Psi}][{\cal D}\Psi]\det[\chi^A,T_B]_{\delta}\,\delta[P_0]\nonumber\\
&\times &\left(\prod_{A=1}^{2}\delta [\chi^A]\right)\,\exp\left\{i\int d^3x
\left[ P_{\mu}{\dot A}^{\mu}+{\dot {\theta}}\,{\cal G}
+i{\bar {\Psi}}\g^{0}{\dot {\Psi}}-{\cal K}
\right]\right\}\,\,,
\eea

\noindent
where

\bea
\label{237}
{\cal K}&=&\frac{m^2}{2}P_iP_i-\frac{m}{2}P_i\e^{ij}A^j+\frac{1}{8}A^iA^i
+\frac{1}{4m^2}F^{ij}F^{ij}-{\bar {\Psi}}\left(i\g^{i}\pa_{i}-M\right)\Psi\nonumber\\
&+&\frac{g}{2}{\bar {\Psi}}\g^i\Psi A^i+mg{\bar {\Psi}}\g^i\Psi\e^{ij}P_j
-\frac{g}{2m}\e_{ij}F^{ij}({\bar {\Psi}}\g^0\Psi)+\frac{g^2}{2}({\bar {\Psi}}\g^0\Psi)^2\nonumber\\
&+&{\cal G}\left(\frac{1}{2}{\cal G}-\frac{1}{2m}\epsilon_{ij}F^{ij}+
g{\bar {\Psi}}\g^0\Psi\right)\,\,,
\eea

\noindent
is the modified Hamiltonian. Once the integration on the variable $\theta$ is carryed 
out, the new first-class constraint ${\cal G}$ can be used to suppress the terms containing 
it, in the above expression. 
Additionaly, the parcial gauge fixing $\chi^1=A^0$ allows the $P_0$ and $A^0$ integrations. The 
lagrangian $A^0_L$ variable is introduced by the Gauss-Law (${\cal G}$) exponentiation. Finally, 
when the integrations on the momentum variables $P_i$ are performed, we arrive at  
  
\be
\label{245}
{\cal W}_{\chi}={\cal N}\,\int\,
[\prod_{\mu=0}^2 {\cal D}A^{\mu}]\,
[{\cal D}\bar{\Psi}][{\cal D}\Psi]
\det[\chi^2,T_2]_{\delta}\delta [\chi^2]
\exp\left[i\int d^3x\,{\cal L}^{MCS}_F\right]\,\,,
\ee

\noindent
where

\bea
\label{250}
{\cal L}^{MCS}_F&=&-\frac{1}{4 m^2}F_{\mu\nu}F^{\mu\nu}
+\frac{1}{4m} \epsilon^{\mu\nu\rho}F_{\mu\nu}A_{\rho}+\bar{\Psi}(i\g^{\mu}\pa_{\mu}-M)\Psi\nonumber\\
&+&\frac{g}{2m}\epsilon_{\mu\nu\rho}F^{\mu\nu}(\bar{\Psi}\g^{\rho}\Psi)-
\frac{g^2}{2}(\bar{\Psi}\g^{\mu}\Psi)(\bar{\Psi}\g_{\mu}\Psi)\,\,\,.
\eea

\noindent
The nonminimal character of the interaction in this lagrangian, is compatible with the 
idea of gauge invariant fermions introduced in (\ref{230}), and a free Gauss-Law (${\cal G}$).
Since ${\cal L}^{MCS}_F$ derives from ${\cal L}^{SD}_F$ through the BFT embedding 
procedure, they must represent equivalent descriptions of the same physical system.

To test this equivalence beyond the formal level, we proceed analysing the lowest 
order contribution to the electron-electron elastic scattering amplitude (M\"oller scattering) 
associated with (\ref{250}). Within the Hamiltonian framework, the dynamics of the system is 
described by the canonical Hamiltonian

\begin{eqnarray}
\label{255}
H & = & \int d^2z \left[ \frac {m^2}{2} P_iP_i\,
-\,\frac{m}{2}P_i \epsilon^{ik} A^k
+\frac {1}{8} A^k A^k + \frac {1}{4m^2}F^{ij} F^{ij}
\right. \nonumber \\ &-&\left.\frac{i}{2}\,{\bar {\Psi}}\,\g^k\,\cdot\pa_k\Psi\,+\,\frac{i}{2}\,
(\pa_k{\bar {\Psi}})\,\g^k\,\cdot\Psi\,-\,M\,{\bar {\Psi}}\,\cdot\Psi\,
+\,\frac{g}{2}({\bar {\Psi}}\g^k\Psi) A^k\right. \nonumber\\ 
&-& \left.\frac{g}{2m} \epsilon_{ij}F^{ij}({\bar {\Psi}}\g^0\Psi)
+\frac{g^2}{2}({\bar {\Psi}}\g^0\Psi)({\bar {\Psi}}\g^0\Psi)
-mg P_i \epsilon^{ik} ({\bar {\Psi}}\g^k\Psi)\right],
\end{eqnarray}

\noindent
the primary constraint $P_0\approx 0$, and the secondary constraint 
${\cal G}\approx 0$. They satisfy a first-class algebra. Once the Coulomb gauge conditions are 
introduced $A^0\approx 0$, $\pa^iA^i\approx 0$, the quantization is performed by means of the 
Dirac bracket quantization procedure \cite{Di1}, which states that the equal time commutation 
rules are abstracted by the corresponding Dirac brackets. Then the constraints, used as 
strong identityes, define a reduced phase-space which is expressed enterly in terms of the fermion 
variables $\Psi$ and ${\bar {\Psi}}$, and the transversal components $P_i^T$, $A^i_T$. In terms of 
this variables, the only nonvanishing equal time (anti)commutators are found to be

\bml
\begin{eqnarray}
\label{260}
\left\{\Psi(x^{0}, \vec{x})\,,\,{\bar {\Psi}}(x^{0}, \vec{y})\right\} & = & \g^0\,
\delta(\vec{x}-\vec{y})\,\,\,,\label{mlett:a260}\\
\left[A^i_T(x^0,\vec{x})\,,\,P^T_j(x^0,\vec{y})\right] & = & i\,\left(\delta^{ij}-\frac{\pa^i_x\pa^j_x}
{\nabla^2_x}\right)\delta(\vec{x}-\vec{y})\,\,\,,\label{mlett:b260}
\end{eqnarray}
\eml

\noindent
while the Hamiltonian splits into a sum of a free part

\be
\label{265}
H_0=\int d^2z\left[\frac {m^2}{2}P_j^T P^T_j
-\frac {m}{2} P_i^T\epsilon^{ik} A^{k}_T+\frac {1}{2} A^k_T A^k_T+\frac {1}{4m^2}F^{ij}_{T}F^{ij}_T
-{\bar {\Psi}}\left(i\g^i\pa_i-M\right){\Psi} \right],
\ee

\noindent
plus an interacting part

\be
\label{270}
H_I=\int d^2z\left[g {\bar {\Psi}} \cdot \g^i\Psi B^i+
- \frac{g}{2m} \epsilon_{ij}F^{ij}({\bar {\Psi}}\cdot\g^0\Psi)
+\frac {g^2}{2}({\bar {\Psi}}\cdot\g^0\Psi)^2\right]\,\,\,,
\ee

\noindent
where $B^i$ is the combination

\be
\label{275}
B^i\,=\,A^i_T\,+\,m\epsilon^{ij}\,P_j^T\,\,.
\ee

\noindent
The dot ($\cdot$) signalizes for the symmetrysation prescription ($\p_a\cdot\bp_b=1/2[\p_a,\bp_b]$) 
adopted in interaction fermion bilinears. From inspection of (\ref{270}), it follows that the 
contributions of order $g^2$ to the above mentioned scattering amplitude, ${\cal R}^{(2)}$, can 
be grouped in five different kinds of terms,

\be
\label{280}
{\cal R}^{(2)}\,=\,\sum_{a=1}^{5}\,{\cal R}^{(2)}_{a}\,\,,
\ee

\noindent
where

\bml
\begin{eqnarray}
\label{285}
R^{(2)}_1 & = & \frac {ig^2}{2}(\gamma ^0)_{ab}(\gamma ^0)_{cd}
\int d^3x\int d^3y
\delta (x^0-y^0)\dxy \nonumber \\
& \times & \langle \Phi _f\vert :{\bar {\Psi}} _a (x)
\Psi _b(x){\bar {\Psi}} _c(y) \Psi _d(y):\vert \Phi _i\rangle\,,
\label{mlett:a285} \\
R^{(2)}_2 & = &-\frac{g^2}{2}(\gamma ^i)_{ab}(\gamma ^j)_{cd}
\int d^3x \int d^3y
\langle \Phi _f\vert T\left\{ :{\bar {\Psi}} _a(x)\Psi _b(x)B^i(x): \right.
\nonumber \\ & \times & \left.  :{\bar {\Psi}} _c
(y)\Psi _d (y)B^j(y):\right\}\vert \Phi _i\rangle\,, \label{mlett:b285} \\
R^{(2)}_3 & = & \frac {g^2}{4m} (\gamma ^i)_{ab}(\gamma ^0)_{cd}
\int d^3x \int d^3y
\langle \Phi _f\vert T\left\{ :{\bar {\Psi}} _a(x)\Psi _b(x)B^i(x): \right.
\nonumber  \\
& \times & \left.  :\e^{jk}F^{jk}_T(y){\bar {\Psi}} _c(y) \Psi _d(y):\right\}\vert \Phi _i\rangle\,, 
\label{mlett:c285} \\
R^{(2)}_4 & = & \frac {g^2}{4m} (\gamma ^0)_{ab}(\gamma ^i)_{cd}
\int d^3x \int d^3y
\langle \Phi _f\vert T\left\{ :\e^{jk}F^{jk}_T(x){\bar {\Psi}} _c(x) \Psi _d(x):
 \right.
\nonumber  \\
& \times & \left.  :{\bar {\Psi}} _a(y)\Psi _b(y)B^i(y):\right\}\vert \Phi _i\rangle\,, 
\label{mlett:d285} \\
R^{(2)}_5 & = & -\frac {g^2}{8m^2}\epsilon ^{kl} \epsilon ^{jm}
(\gamma ^0)_{ab} (\gamma ^0)_{cd} \int d^3x \int d^3y
\langle \Phi _f\vert T\left\{ :\e_{kl}F^{kl}(x){\bar {\Psi}} _a(x)\Psi _b(x): \right.
\nonumber \\ & \times & \left. :\e_{jm}F^{jm}(y)
A^m_T(y)\bar\Psi _c(y)\Psi _d(y):\right\} \vert \Phi _i\rangle\,. \label{mlett:e285}
\end{eqnarray}
\eml

In this equations, $T$ is the chronological ordering product while $\vert \Phi_i\rangle$ and 
$\vert \Phi_f\rangle$ denote the initial and final states of the process, respectively. In 
the present case, they are two-electron states. Fermion states obeying the free Dirac equation 
in $(2+1)$-dimensions were explicitly constructed in Ref.\cite{Gi2}, where the notation 
$v^{(-)}(\bf{p})$(${\bar {v}}^{(+)}(\bf{p})$) was employed to designate the two-component 
spinor describing a free electron of two-momentum ${\bf{p}}$, energy $p^0=+({\bf{p}}^2+M^2)^{1/2}$, 
and spin $s=M/|M|$ in the initial (final) state. The plane wave expansion of the free fermionic 
operators $\Psi$ and ${\bar {\Psi}}$ in terms of these spinors and the corresponding 
creation and annihilation operators goes as usual.

In terms of the initial $(p_1,p_2)$ and final momenta $(p^{\prime}_1,p^{\prime}_2)$, the partial 
amplitudes in (\ref{285}) are found to read

\bml
\begin{eqnarray}
\label{290}
R^{(2)}_{1} & = & -\frac{g^2}{2(2\pi)^2}\delta ^{(3)}(p_1'+p_2'-p_1-p_2)
\nonumber \\
& \times & \left\{ [\bar v^{(+)}(\vec p_1^{\,\prime}) \gamma ^0 v^{(-)}(\vec p_1)]
[ \bar v^{(+)}(\vec p_2^{\,\prime})\gamma ^0 v^{(-)}(\vec p_2)]\,-\, p_1^{\,\prime}
\leftrightarrow p_2^{\,\prime} \right\}, \label{mlett:a290} \\
R^{(2)}_{2} & = & -\frac {g^2}{2(2\pi)^2}\delta^{(3)}(p_1'+p_2'-p_1-p_2)
\nonumber \\
& \times & \left\{ [\bar v^{(+)} (\vec{p}_1^{\,\prime})
\gamma ^l v^{(-)}(\vec p_1)][\bar v^{(+)}(\vec{p}_2^{\,\prime})
\gamma^j v^{(-)}(\vec p_2)] D^{lj}(k)-  p_1^{\,\prime}
\leftrightarrow p_2^{\,\prime} \right\}, \label{mlett:b290} \\
R^{(2)}_{3} & = & -\frac{g^2}{2(2\pi)^2}\delta^{(3)}(p_1'+p_2'-p_1-p_2)
\nonumber \\
& \times & \left\{ [\bar v^{(+)}(\vec p_1^{\,\prime}) \gamma ^l v^{(-)}(\vec p_1)]
[\bar v^{(+)}(\vec p_2^{\,\prime}) \gamma ^0 v^{(-)}(\vec p_2)]
\Gamma ^l(k)\,- \,p_1^{\,\prime} \leftrightarrow p_2^ {\,\prime} \right\},
\label{mlett:c290} \\
R^{(2)}_{4} & = & -\frac{g^2}{2(2\pi)^2}\delta^{(3)}(p_1'+p_2'-p_1-p_2)
\nonumber \\
& \times & \left\{ [\bar v^{(+)}(\vec p_1^{\,\prime}) \gamma ^0 v^{(-)}(\vec p_1)]
[\bar v^{(+)}(\vec p_2^{\,\prime}) \gamma ^l v^{(-)}(\vec p_2)]
\Gamma ^l(-k)\,- \,p_1^{\,\prime} \leftrightarrow p_2^ {\,\prime} \right\},
\label{mlett:d290} \\
R^{(2)}_{5} & = & -\frac{g^2}{2(2\pi)^2}\delta ^{(3)}(p_1'+p_2'-p_1-p_2)
\nonumber \\
& \times & \left\{[\bar v^{(+)} (\vec p_1^{\,\prime})\gamma ^0 v^{(-)}(\vec p_1)]
[\bar v^{(+)}(\vec p_2^{\,\prime})\gamma ^0 v^{(-)}(\vec p_2)]
\Lambda(k)\,- \,\,p_1^{\,\prime} \leftrightarrow p_2^ {\,\prime} \right\},
\label{mlett:e290}
\end{eqnarray}
\eml

\noindent
where

\bml
\bea
\label{295}
D^{ij}(k)&=&\frac{i}{k^2-m^2+i\ve}\left(m^2\delta^{ij}+k^ik^j-im\e^{ij}k^0\right)\,,
\label{mlett:a295}\\
\Gamma^{i}(k)&=&\frac{i}{k^2-m^2+i\ve}\left(k^ik^0+im\e^{in}k^n\right)\,,
\label{mlett:b295}\\
\Lambda(k)&=&\frac{i}{k^2-m^2+i\ve} |\bf{k}|^2\,,
\label{mlett:c295}\,\,.
\eea
\eml

\noindent
and

\begin{equation}
\label{360}
k \equiv  p_1'-p_1\,=\,-(p_2'-p_2),\nonumber
\end{equation}

\noindent
is the momentum transfer. Combining the above parcial amplitudes in (\ref{280}), the 
M\"oller scattering amplitude obtained from (\ref{250}) is given by

\begin{eqnarray}
\label{365}
R^{(2)}\,& = &\,\left( -\frac{g^2}{2(2\pi)^2}\right) \delta^{(3)}(p_1'+p_2'-p_1-p_2)
\nonumber \\
& \times & \left\{ [\bar v^{(+)}(\vec p_1^{\,\prime})
\gamma^{\mu} v^{(-)}(\vec p_1)]\,D_{\mu \nu}(k)\,[\bar v^{(+)}(\vec p_2^{\,\prime})
\gamma^{\nu} v^{(-)}(\vec p_2)] 
- p_1^{\,\prime} \leftrightarrow p_2^ {\,\prime} \right\}\,,
\end{eqnarray}

\noindent
where

\begin{equation}
\label{370}
D_{\mu \nu}(k) = -\frac{i}{k^2-m^2+i\ve }\left( g_{\mu \nu}\,-\frac{k_{\mu}k_{\nu}}{m^2}+\,
im\epsilon _{\mu \nu \rho}\frac{k^{\rho}}{m^2}
\right)\,\,.
\end{equation}

\noindent
The above result is exactlly the one obtained for the same calculation in the SD model minimally 
coupled to Dirac fermions\cite{Gi3}. The result shows that, although the time ordered 
product of the free MCS field is represented by a nonlocal expression\cite{Fleck1}, the particular 
form of the interaction in (\ref{250}) is enough to ensures that the effective 
propagator $D_{\mu \nu}$ which emerge from (\ref{365}) is in fact local an coincides with the 
propagator of the free SD field. 

\section{The SD model coupled to a charged scalar field}
\label{sec:level3}

The dynamics of the SD model minimally coupled to a charged scalar field is 
described by the lagrangian density

\be
\label{1000}
{\cal L}^{SD}_S=- \frac{1}{2m} \epsilon^{\mu\nu\rho}
(\partial_{\mu} f_{\nu})
f_{\rho}+
\frac{1}{2}\,f^{\mu}f_{\mu}+(D^f_{\mu}\phi)^{\ast}(D^{f\mu}\phi)-M^2\phi^{\ast}\phi\,\,\,.
\ee

\noindent
In phase-space, the system is characterized by the primary constraints

\bml
\bea
\label{1005}
&&T_0^{(0)}=\pi_0\,\approx\,0\,\,\,, \label{mlett:a1005} \\
&&T_i^{(0)}=\pi_i
+\frac{1}{2m} \epsilon_{ij} f^j\approx 0\,\,,i=1,2\,\,\,,
\label{mlett:b1005}
\eea
\eml
\noindent
the secondary constraint,

\be
\label{1010}
T_3^{(0)}=\frac{1}{m}\left(f^0
-\frac{1}{m}\epsilon_{ij}\partial^if^j\right)
-\frac{ig}{m}\left(\pi_{\phi}\phi-\pi_{\phi^{\ast}}\phi^{\ast}\right)\approx\,0\,\,\,,
\ee

\noindent
and the canonical hamiltonian

\be
\label{1015}
H^{(0)}=\int d^2x\,\left[\frac{1}{2}f^0f^0+\frac{1}{2}f^if^i
+\pi_{\phi}\pi_{\phi^{\ast}}+(D^f_{i}\phi)^{\ast}(D^f_{i}\phi)+M^2\phi^{\ast}\phi
-\,m f^0T_3^{(0)}\right]\,\,\,.
\ee

Now, we only mention the steps we done here, which are common to the previous case. First, we 
define Dirac $\delta$-brackets with respect to the 
constraints (\ref{mlett:b1005}), so that within the $\delta$-bracket algebra, they holds as 
strong identities and can be used to eliminate the momenta $\pi_i$ from the game. The nonvanishing
$\delta$-brackets are: 

\bml
\bea
\label{1020}
\left [f^i(x^0,\vx)\,,\,f^j(x^0,\vy)\right]_{\delta}&=&-m\epsilon^{ij}\dxy\,\,,
\label{mlett:a1020}\\
\left [\phi(x^0,\vx)\,,\,\pi_{\phi}(x^0,\vy)\,\right]_{\delta}& = &\dxy\,\,\,,
\label{mlett:b1020}\\
\left [\phi^*(x^0,\vx),\pi_{\phi^*}(x^0,\vy)\right]_{\delta}& = &\dxy\,\,\,.
\label{mlett:c1020}
\eea
\eml

\noindent
The BFT embedding procedure is then applyed to the remining system, defined by the 
constraints (\ref{mlett:a1005}), (\ref{1010}) and the Hamiltonian (\ref{1015}). To 
this end, a canonical pair of phase-space variables 
($[\theta,\pi_{\theta}]_{\delta}=1$) is introduced. In terms of them, the BFT embedding 
procedure gives

\bml
\bea
\label{1025}
&&T_0^{(0)} \rightarrow T_1 \,=\,\pi_0+\theta \approx 0\,\,\,,\label{mlett:a1025}\\
&&T_3^{(0)} \rightarrow T_2 \,=\,\frac{1}{m}
(f^0+\pi_{\theta})-\frac{1}{m^2}\e_{ij}\pa^i f^j\,-\,\frac{ig}{m}
(\pi_{\phi}{\phi}-\pi_{\phi^*}{\phi}^*) \,\approx\,0\,\,\,,
\label{mlett:b1025}\\
&&H^{(0)} \rightarrow  H=\int d^2x\left[\frac{1}{2}(f^0+\pi_{\theta})^2+
\frac{1}{2}(f^i+\pa^i\theta)^2+
(D^f_i\phi)^*(D^f_i\phi)+M^2\phi^*\phi\right]\,\,\,.\label{mlett:c1025}
\eea
\eml

\noindent
The new constraints and the new Hamiltonian satisfy a stronglly involutive algebra, and define 
in this way, a first-class theory. The unitarizing Hamiltonian ($H_U$) and the unitarizing action
($A_U$) can be constructed in a form closer to that of the previous section. Once the ghost 
variables are integrated out, the resulting Green function generating functional has 
the form

\bea
\label{1030}
{\cal W}_{\chi}&=&{\cal N}\int [{\cal D}f^0][{\cal D}\pi_0]
[{\cal D}f^{i}][{\cal D}\pi_{\theta}][{\cal D}\theta]
[{\cal D}\phi][{\cal D}\pi_{\phi}][{\cal D}\phi^*][{\cal D}\pi_{\phi^*}]\det[\chi^A,T_B]_{\delta}
\left(\prod_{A=1}^{2}\delta [\chi^A]\right)\nonumber\\
&\times &
\left(\prod_{A=1}^{2}\delta [T_A]\right)
\exp\left[i\int d^3x
\left( \pi_0{\dot f}^0+\frac{1}{2m}f^i\e_{ij}{\dot f}^{j}
+\pi_{\theta}{\dot {\theta}}+\pi_{\phi}{\dot {\phi}}+\pi_{\phi^*}{\dot {\phi}}^*
-{\cal H}\right)\right],
\eea

\noindent
where, ${\cal H}$ is the Hamiltonian density corresponding to the involutive 
Hamiltonian (\ref{mlett:c1025}). Furthermore, under an infinitesimal 
supertransformation generated by the BRST charge ($\Omega$), the phase-space 
variables present in (\ref{1030}), change as follows

\bml
\label{1035}
\bea
\delta\,f^0\,& \equiv &\,[\,f^0\,,\,\Omega\,]_{\delta}\,\ve\,=\,\ve^1
\,\,\,,\label{mlett:a1035}\\
\delta\,f^i\,& \equiv &\,[\,f^i\,,\,\Omega\,]_{\delta}\,\ve\,=\,-\frac{1}{m}\,
\pa^i\ve^2\,\,\,,\label{mlett:b1035}\\
\delta\,\theta\,& \equiv &\,[\,\theta\,,\,\Omega\,]_{\delta}\,\ve\,=\,
\frac{1}{m}\ve^2\,\,\,,\label{mlett:c1035}\\
\delta\,\pi_{0}\,& \equiv &\,[\,\pi_{0}\,,\,\Omega\,]_{\delta}\,\ve\,
=\,-\frac{1}{m}\,\ve^2\,\,\,,\label{mlett:d1035}\\
\delta\,\pi_{\theta}\,& \equiv &\,[\,\pi_{\theta}\,,\,\Omega\,]_{\delta}\,\ve\,
=\,-\,\ve^1\,\,\,,\label{mlett:e1035}\\
\delta\,\phi\,& \equiv &\,[\,\phi\,,\,\Omega\,]_{\delta}\,\ve\,
=\,-\frac{ig}{m}\,\phi\,\ve^2\,\,\,,\label{mlett:f1035}\\
\delta\,\phi^*\,& \equiv &\,[\,\phi^*\,,\,\Omega\,]_{\delta}\,\ve\,
=\,+\frac{ig}{m}\,\phi^*\,\ve^2\,\,\,,\label{mlett:g1035}\\
\delta\,\pi_{\phi}\,& \equiv &\,[\,\pi_{\phi}\,,\,\Omega\,]_{\delta}\,\ve\,
=\,+\frac{ig}{m}\,\pi_{\phi}\,\ve^2\,\,\,,\label{mlett:h1035}\\
\delta\,\pi_{\phi^*}\,& \equiv &\,[\,\pi_{\phi^*}\,,\,\Omega\,]_{\delta}\,\ve\,
=\,-\frac{ig}{m}\,\pi_{\phi^*}\,\ve^2\,\,\,.\label{mlett:i1035}
\eea
\eml

\noindent
The above transformation laws shows that the first-class Hamiltonian (see Eq.(\ref{mlett:c1025}))
cannot be enterly expressed in terms of the gauge invariant combinations 

\bml
\bea
\label{1050}
Y^{0}&=&f^0+\pi_{\theta}\,\,,\label{mlett:a1050}\\
Y^{i}&=&f^i+\pa^i\theta\,\,.\label{mlett:b1050}
\eea
\eml

\noindent
The interaction is still madiated by the gauge dependent field $f^i$. Finally, the 
auxiliary conditions $\chi^1=\pi_{\theta}$ and $\chi^2=\theta$, restores the original 
second-class model and define therefore, the unitary gauge.

Now, as in section 2, we turn to look for alternative formulations for the first-class 
theory. We start by decomposing the basic $f^i$ field in canonical pairs 
($[A^i,P_j]_{\delta}=\delta^i_j$) as 

\be
\label{1055}
f^i\,=\,\frac{1}{2}A^i\,+\,m\epsilon^{ij}\,P_j\,\,\,,
\ee

\noindent
to write the modified generating functional,

\bea
\label{1060}
& &{\cal W}_{\chi}={\cal N}\int [{\cal D}f^0][{\cal D}\pi_0]
[{\cal D}A^{i}][{\cal D}P_{i}][{\cal D}\theta][{\cal D}\pi_{\theta}]
[{\cal D}\phi][{\cal D}\pi_{\phi}][{\cal D}\phi^*][{\cal D}\pi_{\phi^*}]
\det[\chi^A,T_B]_{\delta}\nonumber\\
&\times &
\left(\prod_{A=1}^{2}\delta [\chi^A]\right)
\left(\prod_{A=1}^{2}\delta [T_A]\right)
\exp\left[i\int d^3x
\left( \pi_0{\dot f}^0+P_i{\dot A}^i+\pi_{\theta}{\dot {\theta}}
+\pi_{\phi}{\dot {\phi}}+\pi_{\phi^*}{\dot {\phi}}^* 
-{\cal H}\right)\right].
\eea

\noindent
The equivalence of the above Green function generating functional with (\ref{1030}) is secured 
by the same line of reasoning presented in section 2. The first-class constraint $T_2$, rewriten
in terms of the canonical variables $A^i$ and $P_i$ is

\be
\label{1065}
mT_2\,=\,Y^0+{\cal G}-\frac{1}{2m}\epsilon_{ij}F^{ij}-ig(\pi_{\phi}{\phi}-\pi_{\phi^*}{\phi}^*)
\,\approx\,0\,\,\,,
\ee

\noindent
where $F^{ij}=\pa^iA^j-\pa^jA^i$ and ${\cal G}$ is given by
 
\be
\label{1070}
{\cal G}\,=\,\pa^k P_k\,+\,\frac{1}{2m}\,\e_{kl}\,\pa^k A^l\,\,\,.
\ee

\noindent
We turn next, to Eq.(\ref{1060}), to perform the canonical transformation:

\bml
\bea
\label{1080}
&& f^0\,\rightarrow\,A^{\prime 0}\,=\,f^0\,\,\,,\label{mlett:a1080}\\
&& \pi_0\,\rightarrow\,P^{\prime}_0\,=\,\pi_0\,+\,\theta \,\approx 0\,\,\,,\label{mlett:b1080}\\
&& A^i\,\rightarrow\,A^{\prime i}\,=\,A^i\,\,\,,\label{mlett:c1080}\\
&& P_i\,\rightarrow\,P^{\prime}_i\,=\,P_i\,-\,\frac{1}{m}\,\epsilon_{ij}\,\pa^j\theta
\,\,\,,\label{mlett:d1080}\\
&& \theta\,\rightarrow\,\theta^{\prime}\,=\,\theta\,\,\,,\label{mlett:e1080}\\
&& \pi_{\theta}\,\rightarrow\,\pi_{\theta}^{\prime}\,=\,\pi_{\theta}\,
-\,\frac{1}{2m}\,\e_{ij}\,F^{ij}\,+\,\,f^0\,\,\,,\label{mlett:f1080}\\
&& \phi\,\rightarrow \phi^{\prime}\,=\,\phi\,\,,\label{mlett:g1080}\\
&& \phi^*\,\rightarrow \phi^{\prime *}\,=\,\phi^*\,\,,\label{mlett:h1080}\\
&& \pi_{\phi}\,\rightarrow \pi_{\phi}^{\prime}\,=\,\pi_{\phi}\,\,,\label{mlett:i1080}\\
&& \pi_{\phi^*}\,\rightarrow \pi_{\phi^*}^{\prime}\,=\,\pi_{\phi^*}\,\,.\label{mlett:j1080}
\eea
\eml

\noindent
In the following we omit the``$\prime$''on the new variables. In terms of them, the first-class 
quantities (\ref{mlett:b1050}) read simply $Y^i=1/2A^i+m\e^{ij}P_j$ and, after the $\pi_{\theta}$
integration, the generating functional assumes the form,

\bea
\label{1090}
& &{\cal W}_{\chi}={\cal N}\int
[\prod_{\mu=0}^2 {\cal D}A^{\mu}]
[\prod_{\mu=0}^2 {\cal D}P_{\mu}]
[{\cal D}\theta][{\cal D}\phi][{\cal D}\pi_{\phi}]
[{\cal D}\phi^*][{\cal D}\pi_{\phi^*}]\det[\chi^A,T_B]_{\delta}\delta[P_0]\nonumber\\ 
&\times &
\left(\prod_{A=1}^{2}\delta [\chi^A]\right)
\exp\left\{i\int d^3x
\left[ P_{\mu}{\dot A}^{\mu}+{\dot {\theta}}{\cal G}+
\pi_{\phi}\left( {\dot {\phi}}+ig{\dot {\theta}}\phi\right)
+\pi_{\phi^*}\left( {\dot {\phi}^*}-ig{\dot {\theta}}\phi^*\right)
\right.\right.\nonumber\\
&-&\left.\left.\pi_{\phi}\pi_{\phi^*}-\frac{m^2}{2}P_iP_i+\frac{m}{2}P_i\e^{ij}A^j-\frac{1}{8}A^iA^i
-\frac{1}{4m^2}F^{ij}F^{ij}
\right.\right.\nonumber\\
&-&\left.\left.\left( {\pa_i\phi^*+ig(Y_i-\pa_i\theta)\phi^*}\right)
\left( {\pa_i\phi-ig(Y_i-\pa_i\theta)\phi}\right)-\frac{ig}{2m}\e_{ij}F^{ij} (\pi_{\phi}{\phi}
-\pi_{\phi^*}{\phi}^*)
\right.\right.\nonumber\\
&+&\left.\left.\frac{g^2}{2}(\pi_{\phi}{\phi}-\pi_{\phi^*}{\phi}^*)^2
-{\cal G}\left(\frac{1}{2}{\cal G}-\frac{1}{2m}\epsilon_{ij}F^{ij}-
ig(\pi_{\phi}{\phi}-\pi_{\phi^*}{\phi}^*)\right)\right]\right\}\,\,.
\eea
      
\noindent
The next step, is to define the gauge invariant scalars

\bml
\bea
\label{1100}
\Phi &=& \phi\,\exp{(+ig\theta)}\,\,,\label{mlett:a1100}\\
\Phi^* &=& \phi^*\,\exp{(-ig\theta)}\,\,,\label{mlett:b1100}\\
\pi_{\Phi} &=& \pi_{\phi}\,\exp{(-ig\theta)}\,\,,\label{mlett:c1100}\\
\pi_{\Phi} &=& \pi_{\phi}\,\exp{(+ig\theta)}\,\,.\label{mlett:d1100}
\eea
\eml

\noindent
in terms of which all terms containing $\theta$, exepting the one involving ${\cal G}$, drop out 
to produce, after the $\theta$ integration, the generating functional

\bea
\label{1110}
& &{\cal W}_{\chi}={\cal N}\int
[\prod_{\mu=0}^2 {\cal D}A^{\mu}]
[\prod_{\mu=0}^2 {\cal D}P_{\mu}]
[{\cal D}\theta][{\cal D}\phi][{\cal D}\pi_{\phi}]
[{\cal D}\phi^*][{\cal D}\pi_{\phi^*}]\det[\chi^A,T_B]_{\delta}\nonumber\\ 
&\times &\delta[{\cal G}]\delta[P_0]\left(\prod_{A=1}^{2}\delta [\chi^A]\right)
\exp\left\{i\int d^3x
\left[ P_{\mu}{\dot A}^{\mu}+
\pi_{\Phi}{\dot {\Phi}}+\pi_{\Phi^*}{\dot {\Phi}^*}-{\cal K}\right]\right\}\,\,.
\eea

\noindent
Here, ${\cal K}$ is the modifyed Hamiltonian

\bea
\label{1120}
{\cal K} &=&\pi_{\Phi}\pi_{\Phi^*}+\frac{m^2\alpha}{2}P_iP_i
-\frac{m\alpha}{2}P_i\e^{ij}A^j
+\frac{\alpha}{8}A^iA^i
+\frac{1}{4m^2}F^{ij}F^{ij}\nonumber\\
&+& (\pa_i\Phi^*)(\pa_i\Phi)+M^2\Phi^*\Phi-\frac{ig}{2}(\Phi^{*}{\qv{\pa}}_{i}\Phi)A^i
+img P_i\e^{ij}(\Phi^{*}{\qv{\pa}}_{j}\Phi)\nonumber\\
&+&\frac{ig}{2m}\e_{ij}F^{ij} (\pi_{\Phi}{\Phi}-\pi_{\Phi^*}{\Phi}^*)
-\frac{g^2}{2}(\pi_{\Phi}{\Phi}-\pi_{\Phi^*}{\Phi}^*)^2\,\,,
\eea

\noindent
and $\alpha$ is the factor

\be
\label{1130}
\alpha\,=\,(1+2g^2|\Phi|^2)\,\,.\nonumber
\ee

\noindent
The expression (\ref{1110}), constructed from the original second-class model 
through the BFT embedding procedure, the canonical transformation (\ref{1080}) and the gauge 
invariant scalars (\ref{1100}), characterizes a particular phase-space realization for the Green 
function generating functional associated to the first-class theory.

We show now, that the non polinomial lagrangian density 

\bea
\label{1140}
{\cal L}^{MCS}_{S}&=&-\frac{1}{4 m^2\alpha}F_{\mu\nu}F^{\mu\nu}
+\frac{1}{4m} \epsilon^{\mu\nu\rho}F_{\mu\nu}A_{\rho}
+(\pa_{\mu}{\Phi}^{*})(\pa^{\mu}\Phi)-M^2\Phi^{*}\Phi\nonumber\\
&+&\frac{ig}{2m\alpha}\epsilon^{\mu\nu\rho}F_{\mu\nu}(\Phi^{*}{\qv{\pa}}_{\rho}\Phi)
+\frac{g^2}{2\alpha}(\Phi^{*}{\qv{\pa}}_{\mu}\Phi)(\Phi^{*}{\qv{\pa}}^{\mu}\Phi)\,\,\,,
\eea

\noindent
leads to the same particular phase-space realization for the Green function 
generating functional. Concerning the above mentioned model, we start by defining the 
momentum variables 

\bml
\bea
\label{1150}
P_0 & \approx & 0\,\,\,,\label{mlett:a1150}\\
P_i &=&\frac{1}{m^2\alpha}F^{0i}+\frac{1}{2m}\e_{ij}A^j
-\frac{ig}{m\alpha}\e_{ij}(\Phi^{*}{\qv{\pa}}_{j}\Phi)\,\,\,,\label{mlett:b1150}\\
\pi_{\Phi}&=&{\dot {\Phi}}^*+\frac{ig}{2m\alpha}(\e{\cdot}F)\Phi^*
+\frac{g^2}{\alpha}(\Phi^{*}{\qv{\pa}}_{j}\Phi)^2\Phi^*\,\,\,,\label{mlett:c1150}\\
\pi_{\Phi^*}&=&{\dot {\Phi}}-\frac{ig}{2m\alpha}(\e{\cdot}F)\Phi
-\frac{g^2}{\alpha}(\Phi^{*}{\qv{\pa}}_{j}\Phi)^2\Phi\,\,\,,\label{mlett:d1150}
\eea
\eml

\noindent
canonically conjugate to the fields $A^0$, $A^i$, ${\Phi}$ and ${\Phi}^*$ respectively. The 
expressions (\ref{mlett:c1150}) and (\ref{mlett:d1150}) can be combined to produce the 
kinetic term

\bea
\label{1160}
\pi_{\Phi}{\dot {\Phi}}+\pi_{\Phi^*}{\dot {\Phi}}^* &=& 2\pi_{\Phi}\pi_{\Phi^*}
+\frac{ig}{2m\alpha}(\e{\cdot}F)(\pi_{\Phi}{\Phi}-\pi_{\Phi^*}{\Phi}^*)
\nonumber\\ &+& \frac{g^2}{\alpha}(\Phi^{*}{\qv{\pa}}_{0}\Phi)(\pi_{\Phi}{\Phi}-\pi_{\Phi^*}{\Phi}^*)
\,\,\,,
\eea

\noindent
and the electric charge density

\be
\label{1170}
\pi_{\Phi}{\Phi}-\pi_{\Phi^*}{\Phi}^* \,=\,-\frac{1}{\alpha}\left[(\Phi^{*}{\qv{\pa}}_{0}\Phi)
-\frac{ig}{m}(\e{\cdot}F)|\Phi|^2\right]\,\,\,.
\ee

\noindent
The momenta, thogether with the kinetic term, are used to write the 
canonical Hamiltonian density

\be
\label{1180}
{\cal H}_c\,=\,P_{\mu}{\dot {A}}^{\mu}+\pi_{\Phi}{\dot {\Phi}}
+\pi_{\Phi^*}{\dot {\Phi}}^*-{\cal L}^{MCS}_S\,\,\,,
\ee

\noindent
as

\bea
\label{1190}
{\cal H}_c &=& -A^0\left(\pa^iP_i+\frac{1}{2m}\e_{ij}\pa^iA^j\right)+\pi_{\Phi}\pi_{\Phi^*}
+(\pa_i\Phi^*)(\pa_i\Phi)+M^2\Phi^*\Phi\nonumber\\
&+&\frac {m^2\alpha}{2}P_i P_i
-\frac {m\alpha}{2} P_i\epsilon^{ik} A^{k}+\frac {\alpha}{2} A^kA^k
+\frac {1}{4m^2\alpha}F^{ij}F^{ij}(1-\frac{2g^2}{\alpha}|\Phi|^2)\nonumber\\
&-&\frac{ig}{2m\alpha}(\e{\cdot}F) (\Phi^{*}{\qv{\pa}}_{0}\Phi)(1-\frac{2g^2}{\alpha}|\Phi|^2)
-\frac{g^2}{2\alpha}(\Phi^{*}{\qv{\pa}}_{0}\Phi)^2(1-\frac{2g^2}{\alpha}|\Phi|^2)\nonumber\\
&+&img P_i\e^{ij}(\Phi^{*}{\qv{\pa}}_{j}\Phi)-\frac{ig}{2}(\Phi^{*}{\qv{\pa}}_{i}\Phi)A^i\,\,,
\eea

\noindent
where $1-2g^2|\Phi|^2/\alpha\equiv 1/\alpha$. On the other hand, the charge 
density (\ref{1170}) allows 
us to eliminate the velocity $(\Phi^{*}{\qv{\pa}}_{0}\Phi)$. The 
resulting terms can be grouped to form

\bea
\label{1200}
{\cal H}_c &=& -A^0\left(\pa^iP_i+\frac{1}{2m}\e_{ij}\pa^iA^j\right)+\pi_{\Phi}\pi_{\Phi^*}
+(\pa_i\Phi^*)(\pa_i\Phi)+M^2\Phi^*\Phi\nonumber\\
&+&\frac {m^2\alpha}{2}P_i P_i
-\frac {m\alpha}{2} P_i\epsilon^{ik} A^{k}+\frac {\alpha}{8} A^kA^k
+\frac{1}{4m^2\alpha^2}F{\cdot}F(1+4g^2|\Phi|^2+4g^4|\Phi|^4)\nonumber\\
&+&\frac {ig}{2m\alpha}(\e{\cdot}F)(\pi_{\Phi}{\Phi}-\pi_{\Phi^*}{\Phi}^*)(1+2g^2|\Phi|^2)
-\frac{g^2}{2\alpha^2}(\pi_{\Phi}{\Phi}-\pi_{\Phi^*}{\Phi}^*)^2\alpha^2\nonumber\\
&+&img P_i\e^{ij}(\Phi^{*}{\qv{\pa}}_{j}\Phi)-\frac{ig}{2}(\Phi^{*}{\qv{\pa}}_{i}\Phi)A^i\,\,,
\eea

\noindent
which, once we realize that, $1+4g^2|\Phi|^2+4g^4|\Phi|^4\equiv\alpha^2$, reduces to

\bea
\label{1210}
{\cal H}_c &=& -A^0\left(\pa^iP_i+\frac{1}{2m}\e_{ij}\pa^iA^j\right)+\pi_{\Phi}\pi_{\Phi^*}
+(\pa_i\Phi^*)(\pa_i\Phi)+M^2\Phi^*\Phi\nonumber\\
&+&\frac {m^2\alpha}{2}P_i P_i
-\frac {m\alpha}{2} P_i\epsilon^{ik} A^{k}+\frac {\alpha}{8} A^kA^k
+\frac{1}{4m^2}F^{ij}F^{ij}\nonumber\\
&+&img P_i\e^{ij}(\Phi^{*}{\qv{\pa}}_{j}\Phi)-\frac{ig}{2}(\Phi^{*}{\qv{\pa}}_{i}\Phi)A^i\nonumber\\
&+&\frac {ig}{2m}(\e{\cdot}F)(\pi_{\Phi}{\Phi}-\pi_{\Phi^*}{\Phi}^*)
-\frac{g^2}{2}(\pi_{\Phi}{\Phi}-\pi_{\Phi^*}{\Phi}^*)^2\,\,.
\eea

\noindent
The above expression, for the canonical Hamiltonian associated to the lagrangian density 
(\ref{1140}), coincides with that of (\ref{1120}). Moreover, we see that the time persistance of 
the primary constraint (\ref{mlett:a1150}), produces the free Gauss-law (\ref{1070}). In this way, the 
model we just described, has the same phase-space realization for the Green function 
generating functional as the first-class model constructed through the BFT embedding 
procedure. Therefore, the configuration space verssion for the Green function generating functional 
corresponding to (\ref{1110}) is

\be
\label{1220}
{\cal W}_{\chi}={\cal N}\int
[\prod_{\mu=0}^2 {\cal D}A^{\mu}][{\cal D}\Phi][{\cal D}\Phi^*]
\det[\chi^2,{\cal G}]_{\delta}\delta[{\cal G}]\delta [\chi^2]
\exp\left\{i\int d^3x\,{\cal L}^{MCS}_S\right\}\,\,.
\ee

\noindent
Since ${\cal L}^{MCS}_S$ derives from 
${\cal L}^{SD}_S$ through the BFT embedding procedure, they must be equivalent descriptions of the 
same physical reality.

Concerning the M\"oller scattering amplitude associated to these models, we only mention 
that they coincide. The calculations lead to the following common result for the amplitude

\begin{eqnarray}
\label{1230}
R^{(2)}\,& = &\, -\frac{g^2}{2(2\pi)^2}{\delta^{(3)}(p_1'+p_2'-p_1-p_2)}
\frac {m^2}{\sqrt{{p_1^0}{p_2^0}{p^{\prime 0}_1}{p^{\prime 0}_2} }}
\nonumber \\
& \times & \left\{(p_1^{\mu}+p_1^{\prime\mu})\,D_{\mu \nu}(k)\,
(p_2^{\nu}+p_2^{\prime\nu}) - p_1^{\prime} 
{\leftrightarrow} p_2^{\prime} \right\}\,,
\end{eqnarray}

\noindent
where

\begin{equation}
\label{1240}
D_{\mu \nu}(k) = -\frac{i}{k^2-m^2+i\ve }\left( g_{\mu \nu}\,-\frac{k_{\mu}k_{\nu}}{m^2}+\,
im\epsilon _{\mu \nu \rho}\frac{k^{\rho}}{m^2}
\right)\,\,,
\end{equation}

\noindent
is the free SD propagator.

\section{Concluding remarks}
\label{sec:level4}
In the present work, we have shown through two examples of interacting field theories, that 
the BFT embedding procedure can be viewed as a systematics to generate a set of first-class
theories equivalent to a given second-class one. Our proof of equivalence is not restricted 
to demonstrate the equality between two functional integrals, but also, involves an explicit 
construction of the phase-space variables of one model in terms of those of the other. This 
is the meaning of the canonical transformation implemented in each case. 

It has recently appeared in the literature the BFT embedding of the non Abelian SD 
model\cite{Rothe6}. The generalization for this case, of our technique mounted on canonical 
transformations, is currently under progress.

\section{Acknowledgements}
\label{sec:level5}

To H. O. Girotti, V. O. Rivelles, M. O. C. Gomes
and M. M. Leite, for the helpfull sugestions and discussions. We would
also make a special thank to V. O. Rivelles and Universsidade de
S\~{a}o Paulo for its kind  hospitality, and to FAPESP for the
financial support.

\end{document}